\documentclass[twocolumn,showpacs,preprintnumbers]{revtex4}

\newtheorem{definition}{Definition}

\newcommand{\blam}{{\bar\lambda}}
\newcommand{\De}{{\Delta}}

\newcommand{\vps}{{\varepsilon}}
\newcommand{\bvps}{{\bar\varepsilon}}

\newcommand{\al}{\alpha}
\newcommand{\bal}{{\bar\alpha}}
\newcommand{\be}{\beta}
\newcommand{\bbe}{{\bar\beta}}
\newcommand{\de}{\delta}

\newcommand{\bxi}{{\bar\xi}}
\newcommand{\bX}{{\bar X}}
\newcommand{\del}{{\nabla}}

\newcommand{\bm}{{\bar m}}

\newcommand{\bmu}{{\bar\mu}}

\newcommand{\bpi}{{\bar\pi}}

\newcommand{\bome}{{\bar\omega}}

\newcommand{\bze}{{\bar\zeta}}

\newcommand{\ppt}{{\frac{\partial}{\partial t}}}
\newcommand{\ppr}{{\frac{\partial}{\partial r}}}

\newcommand{\ppz}{{\frac{\partial}{\partial\zeta}}}
\newcommand{\ppbz}{{\frac{\partial}{\partial{\bar\zeta}}}}
\newcommand{\heq}{{\hat =}}
\newcommand{\cL}{{\cal L}}
\newcommand{\cH}{{\cal H}}
\newcommand{\dV}{{\dot V}}
\newcommand{\cA}{{\cal A}}
\newcommand{\cB}{{\cal B}}
\newcommand{\cC}{{\cal C}}
\newcommand{\bean}{\begin{eqnarray}}

\newcommand{\eean}{\end{eqnarray}}

\newcommand{\eq}[1]{Eq. (\ref{#1})}
\newcommand{\meq}[1]{(\ref{#1})}
\newcommand{\tr}{\tilde r}
\newcommand{\ppa}[1]{\left(\frac{\partial}{\partial #1}\right)}

\begin{document}

\title{\bf Tunneling Effect Near Weakly Isolated Horizon }

\author{Xiaoning Wu}
\email{ wuxn@amss.ac.cn}
\affiliation{ Institute of Applied Mathematics,
Academy of Mathematics and System Science, Chinese Academy of
Sciences, Beijing, China, 100080.}

\author{Sijie Gao}
\email{sijie@bnu.edu.cn}
\affiliation{ Department of Physics,
Beijing Normal University, Beijing 100875, China}

\begin{abstract}
The tunneling effect near a weakly isolated horizon (WIH) has been
studied. By applying the null geodesic method of Parikh and
Wilczek and Hamilton-Jacibi method of Angheben et al. to a weakly
isolated horizon, we recover the semiclassical emission rate in
the tunneling process. We show that the tunneling effect exists in
a wide class of spacetimes admitting weakly isolated horizons. The
general thermodynamic nature of WIH is then confirmed.
\end{abstract}

\pacs{04.70.Dy, 03.65.Sq, 04.62.+v}

\maketitle

\section{Introduction}
Since the discovery of the first exact solution of Einstein
equations, studying properties of black hole is always a highlight
of gravitational physics. Laws of black hole mechanics and the
famous Hawking radiation reveal deep connections between classical
general relativity, quantum physics, and statistical mechanics. It
is surprising that all stationary black hole can be described by
several elegant laws. Furthermore, black holes are playing a major
role in relativistic astrophysics, providing mechanisms to fuel
the most powerful engines in the cosmos. In experimental areas,
black holes are also important sources for gravitational wave
detection. Unfortunately, the classical definition of black hole
can not satisfy the requirements of practical research\cite{As04}.
We need a ``quasi-local" definition for black hole. In the last
decade, much work has been done in this
area\cite{Hay94}-\cite{As04}. In contrast to event horizons, the
quasi-local treatment doesn't require the global description of
spacetime. Therefore, black hole mechanical laws can be
generalized to horizons while ADM quantities are replaced by local
defined quantities. A natural question is whether these mechanical
laws have thermal dynamical explanation. For stationary black
holes, the answer is positive due to the discovery of Hawking
radiation\cite{Haw75,GH77}. Generalization to isolated horizons is
not straightforward  because Hawking's method requires the
knowledge of the global geometry in a space-time, not just the
geometry near a horizon.

Recently, a new technique concerning Hawking radiation has been
developed by Wilczek and his
colleagues\cite{Wil95,Par00,Par2,Par1}. By treating Hawing
radiation as a tunneling process near the Schwarzschild horizon
and using WKB approximation, a semi-classical emission rate was
derived as \bean \Gamma\sim e^{\bigtriangleup S_{B-H}}\,,
\label{gsb} \eean where $\bigtriangleup S_{B-H}$ is the change of
the Bekenstein-Hawking entropy. Although this result is not
exactly the thermal spectrum as first discovered by Hawking, it is
consistent with the conservation of energy during the process.
 This method has been successfully extended to some other
types of black holes\cite{ANVZ05,KM06,zhao06}. All these
calculations rely on specific forms of space-time. Since the
tunneling effect only concerns the physics around a horizon, it is
natural to guess that tunneling effect could be a general
phenomenon for all isolated horizons. The isolated horizon theory
developed by Ashtekar {\em et al.} provides a perfect tool to
explore this issue. By investigating the outgoing null geodesics
near the horizon and applying the first law of the WIH, we recover
\eq{gsb} for the WIH. The WIH contains the essential features of a
black hole, including mass, change and angular momentum. Thus, our
work generalizes the result of the tunneling effect from
stationary black holes to a wide class of spacetimes that admit
WIHs.

Angheben et al.\cite{ANVZ05}
 reconsidered this problem using Hamilton-Jacobi ansatz.
 The Hamilton-Jacobi provides an alternative derivation of the tunneling effect.
It is also a local method involving only an infinitesimal region
around the horizon. Similarly, we apply this approach to a WIH,
without the knowledge of the spacetime metric, we find the
emission rate agreews with \eq{gsb}. Therefore, it confirms that
the Hawking-like radiation exists in very general background.

This paper is organized as follows. In section \ref{geometry}, we
first give a brief review of Ashtekar's weakly isolated horizon
theory. Technical details can be found in ref.\cite{As04} and
references therein. In relation to WIH, we introduce the
Bondi-like coordinates and calculate the outgoing null geodesic
using Newman-Panrose formulism. Section \ref{tew} contains the
calculations on the tunneling effect. In section \ref{mone} and
\ref{sec-hj},  Wilczek's null geodesic method and Angheben's
Hamilton-Jacobi method are employed respectively to study the
tunneling effect near a WIH. By using the result in section
\ref{geometry}, we find that both approaches lead to the desired
spectrum \eq{gsb}. In section \ref{sec-coo}, we discuss the role
of coordinates in the calculations. The master equation in the
Parikh-Wilczek method \cite{Par00} apparently depends on the
choice of coordinates. The results in \ref{mone} apparently depend
on the Bondi-like coordinate system chosen in section
\ref{geometry}. In section \ref{sec-coo}, we analyze the freedom
of the coordinates that keeps the results invariant. We show that
what is relevant in the calculation is not the coordinates, but
the vector field approaching the horizon. Finally, we make some
concluding remarks in section \ref{conclusion}.

\section{Geometry of Weakly Isolated Horizon}\label{geometry}
Base on works by Ashtekar and other authors\cite{As99}-\cite{As04},
the weakly isolated horizon is defined by
\begin{definition}

Let $(M, g)$ be a space-time. $\cH$ is a 3-dim null hyper-surface in
$M$ and $l^a$ is the tangent vector filed of the generator of $\cH$.
$\cH$ is said to be a {\bf weakly isolated horizon}(WIH), if

1) $\cH$ has the topology of $S^2\times\Re$,

2) The expansion of the null generator of $\cH$ is zero, i.e.
$\Theta_l=0$ on $\cH$,

3) $T_{ab}v^b$ is future causal for any future causal vector $v^a$
and Einstein equation holds in a neighborhood of $\cH$,

4) $[\cL_l\ ,\ D_a]l^b=0$ on $\cH$, where $D_a$ is the induced
covariant derivative on $\cH$.
\end{definition}
Since we are going to study the null geodesic behavior around the
horizon,  we need non-singular coordinates to describe the
geometry through the horizon. For the Schwarzschild black hole,
Parikh and Wilczek \cite{Par00} chose the Painlev\'e coordinates.
For the WIH, a convenient choice is the Bondi-like coordinates
\cite{PR86}. Such coordinates have
been used to prove the local existence of WIH \cite{Lew00}. The
coordinates are constructed as following: The tangent vector of
null generator of $\cH$ is $l^a$. Another real null vector field
is $n^a$. The foliation of $\cH$ gives us the natural coordinates
$(\theta,\phi)$. Using a parameter $u$ of $l^a$ and Lie dragging
$(\theta,\phi)$ along each generator of $\cH$, we have coordinates
$(u,\theta,\phi)$ on $\cH$. Choose the affine parameter $r$ of
$n^a$ as the forth coordinate, we get the Bondi-like coordinates
$(u,r,\theta,\phi)$ near the horizon. In these coordinates, we
choose the null tetrad as
\begin{eqnarray}
l^a&=&\frac{\partial}{\partial u}+U\frac{\partial}{\partial
r}+X\frac{\partial}{\partial
\zeta}+\bX\frac{\partial}{\partial\bze}\ ,\nonumber\\
n^a&=&\frac{\partial}{\partial r}\ ,\nonumber\\
m^a&=&\omega\ppr+\xi^3\ppz+\xi^4\ppbz\ ,\nonumber\\
\bm^a&=&\bome\ppr+\bxi^3\ppbz+\bxi^4\ppz,\label{tetrad}
\end{eqnarray}
where $U=X=\omega=0$ on $\cH$ (following the notation in
ref.\cite{As04}, equalities restricted to $\cH$ will be denoted by
`` $\heq$ "), $\zeta=e^{i\phi}\cot\frac{\theta}{2} $. In words of
Newman-Penrose formulism, we also require the above tetrad satisfy
following gauge,
\begin{eqnarray}
&&\nu=\gamma=\tau=\al+\bbe-\pi=\mu-\bmu=0,\nonumber\\
&&\vps-\bvps\ \heq\ \kappa\ \heq\ 0, \label{gaugec}
\end{eqnarray}
which means these tetrad vectors are parallelly transported along
$n^a$ in space-time and Lie drag-free along $l^a$ on $\cH$.
Furthermore, the forth requirement in the definition of WIH
implies there exists a one form $\omega_a$ on $\cH$ such that
$D_al^b\heq\ \omega_al^b$ and $\cL_l\omega_a\heq\ 0$. In terms of
the Newman-Penrose formulism, $\omega_a$ can be expressed as
$\omega_a=-(\vps+\bvps)n_a+(\al+\bbe)\bm_a+(\bal+\be)m_a$. The
above equation means $(\vps+\bvps)$ is constant on $\cH$. In such
a space-time, the out-going null geodesic is described by $l^a$.
We have known that $U\heq 0$. Commutative relation of $l^a$ and
$n^a$ tells us that
\begin{eqnarray}
\frac{\partial U}{\partial r}=(\vps+\bvps)-\bpi\bome-\pi\omega,
\end{eqnarray}
which means $\frac{\partial U}{\partial r}\heq(\vps+\bvps)$. Then
the behavior of function $U$ near $\cH$ is
\begin{eqnarray}
U=(\vps+\bvps)r+o(r).
\end{eqnarray}
Based on the discussion in ref.\cite{As04}, not any choice of time
direction can give a Hamiltonian evolution. In other words, only
some suitably chosen time direction can lead to a well-defined
horizon mass. In ref.\cite{As04}, Ashtekar {\em et al.} gave a
canonical way to choose the time direction $t^a$ for a WIH.
Compared with the Schwarzschild case \cite{Par00}, the parameter
of $t^a$ takes the place of the Killing time. Using the tetrad in
\eq{tetrad}, We find the time derivative of $r$ along the
outgoing geodesic
\begin{eqnarray} {\dot
r}=\frac{du}{dt}\frac{dr}{du}=(B_t+o(1))U=B_t(\vps+\bvps)r+o(r).\label{dr}
\end{eqnarray}
By definition, the surface gravity of $\cH$ is
$\kappa_t:=B_tl^a\omega_a=B_t(\vps+\bvps)$. Because
$B_t(\vps+\bvps)$ is constant on $\cH$, the zeroth law of black
hole mechanics is valid for WIH.

With the canonical time direction $t^a$, Ashtekar {\em et
al.}\cite{As04} showed that the following relation holds on $\cH$.
\begin{eqnarray}
\de M^{(t)}_{\cH}=\frac{\kappa_t}{8\pi}\ \de a_{\cH}+\Omega_t\de
J_{\cH},\label{1}
\end{eqnarray}
where $M^{(t)}_{\cH}$ is the horizon mass, $a_{\cH}$ is the area
of the cross section of WIH, $\Omega_t$ is the angular velocity of
the horizon and $J_{\cH}=-\frac{1}{8\pi}\oint_S(\omega_a\psi^a)dS$
is the angular momentum. This is the first law of black hole
mechanics for WIH, which is a generalization of the first law for
stationary black holes.

\section{Tunneling Effect Near Weakly Isolated Horizon}\label{tew}
\subsection{Null geodesic method}\label{mone}
In this section, we consider the tunneling effect near a weakly
isolated horizon. We apply Parikh and Wilczek's semi-classical
method \cite{Par00,Par2,Par1} to a WIH. We shall use the
Bondi-like coordinates introduced in section \ref{geometry}. It
has been demonstrated\cite{Par00} that the WKB approximation is
justified near a horizon. From the WKB formula, the emission rate
$\Gamma$ can be expressed as
\begin{eqnarray}
\Gamma\sim\exp(-2ImS).
\end{eqnarray}
If the positive energy particle goes outwards, from $r_{in}$ to
$r_{out}$, the imaginary part of action is given by
\begin{eqnarray}
Im S&=&Im\int^{r_{out}}_{r_{in}}p_rdr,\nonumber\\
&=&Im\int^{r_{out}}_{r_{in}}\int^{p_r}_0 dp_rdr.
\end{eqnarray}
From the Hamiltonian equation, $p_r$ can be expressed as
\begin{eqnarray}
dp_r=\frac{dH}{\dot r} \label{dpdh}
\end{eqnarray}
With the help of \eq{dpdh} and Eq.(\ref{dr}), the imaginary part
of the action becomes
\begin{eqnarray}
Im S&=&Im\int^{M-\omega}_{M}\int^{r_{out}}_{r_{in}}\frac{dH}{\dot r}\ dr\nonumber\\
&=&Im\int^{M-\omega}_{M}\int^{r_{out}}_{r_{in}}\frac{dr}{B_t(\vps+\bvps)r+o(r)}\ dH\nonumber\\
&=&\pi\int\frac{dH}{B_t(\vps+\bvps)}\quad.
\end{eqnarray}
Following the argument in \cite{Wil95,Par00}, we fix the total
mass of the space-time and vary the black hole mass. In section
\ref{geometry}, the general form of the first law is given by
Eq.(\ref{1}). For the spherically symmetric case ($\Omega_t=0$),
we can express $dH$ in terms of the surface gravity $\kappa_t$ and
the horizon area $a_{\cH}$,
\begin{eqnarray}
dH=-dM^{(t)}_{\cH}=-\frac{\kappa_t}{8\pi}\
da_{\cH}=-\frac{\kappa_t}{2\pi}\ d\left(\frac{a_{\cH}}{4}\right).
\end{eqnarray}
The minus sign in above equation comes from the conservation of
energy. Then we have
\begin{eqnarray}
Im S&=&\pi\int\frac{dH}{B_t(\vps+\bvps)}\nonumber\\
&=&-\pi\int\frac{\kappa_t}{2\pi B_t(\vps+\bvps)}\ d\left(\frac{a_{\cH}}{4}\right)\nonumber\\
&=&-\frac{1}{2}\ \De\left(\frac{a_{\cH}}{4}\right),
\end{eqnarray}
and the emission rate $\Gamma$ is
\begin{eqnarray}
\Gamma\sim\exp(-2ImS)=\exp(\De\frac{a_{\cH}}{4})=\exp(\De
S_{B-H})\quad.\label{spe}
\end{eqnarray}
This gives the thermal spectrum of radiation, which is in
agreement with \eq{gsb}.

For an axial symmetric horizon, the first law
becomes\cite{As01,As04}
\begin{eqnarray}
dM_{\cH}=\frac{\kappa_t}{8\pi}\ da_{\cH}+\Omega_t\
dJ_{\cH}.\label{1law}
\end{eqnarray}
Using the formula in ref.\cite{zhao06} and the first law \meq{1},
we have
\begin{eqnarray}
ImS&=&Im\int(L-P_{\phi}{\dot\phi})dt\nonumber\\
&=&Im\int[P_r-\frac{P_{\phi}{\dot\phi}}{{\dot r}}]dr\nonumber\\
&=&Im\left(\int\int [dP_r-\frac{{\dot\phi}dP_{\phi}}{\dot r}]dr\right)\nonumber\\
&=&Im\left(\int\frac{dH-{\dot\phi}dP_{\phi}}{\dot r}\ dr\right) \nonumber\\
&=&Im\left(\int\frac{-dM_{\cH}+\Omega_tdJ_{\cH}}{\dot r}\ dr\right)\nonumber\\
&=&-\frac{1}{2}\ \De\left(\frac{a_{\cH}}{4}\right)\,, \label{sah}
\end{eqnarray}
where the conservation of angular momentum has been used in the
fifth equality. \eq{sah} gives the same emission rate near a
rotating horizon. Therefore, we have proven that the semiclassical
emission rate in form of \eq{gsb} holds for a general WIH.

\subsection{Hamilton-Jacobi method}\label{sec-hj}
In this section, we use the Hamilton-Jacobi method to re-calculate
the Hawking radiation near a WIH. This method has been used in
stationary space-times\cite{ANVZ05,KM06}.  We need first modify
Bondi-like coordinates $(t,r,\theta,\phi)$. In general, we can
always write $t^a$ as $t^a\heq
B_tl^a-X^A_0\left(\frac{\partial}{\partial x^A}\right)^a$, where
$A,B=3,4$, $X_0^3=B_tX|_{\cH}$, $X_0^4=B_t\bX|_{\cH}$,
$x^3=\zeta$, $x^4=\bze$. The modified Bondi coordinates are
constructed in following way: Instead of using $u$, we use
parameter of $t^a$ as the time coordinate; the coordinates
$(\theta,\phi)$ are still obtained from a foliation of $\cH$ and
extended to $\cH$ by Lie-dragged along $t^a$; coordinate $r$ is
chosen such that $n^a=f\left(\frac{\partial}{\partial
r}\right)^a$. Under these coordinates, the only different gauge is
that $X$ is constant on $\cH$. In these coordinates, we write the
tetrad as
\begin{eqnarray}
l^a&=&\frac{1}{B_t}\ppt+U\ppr+X\ppz+\bX\ppbz,\nonumber\\
n^a&=&f\ppr,\nonumber\\
m^a&=&\omega\ppr+\xi^3\ppz+\xi^4\ppbz,\nonumber\\
\bm^a&=&\bome\ppr+\bxi^3\ppbz+\bxi^4\ppz.\label{mtetrad}
\end{eqnarray}
where $B_tX\heq X_0$, $f\heq 1$, $\omega\heq 0$, $U\heq 0$. Then
the metric $g^{\mu\nu}$ takes the form
\begin{eqnarray}
\left(\begin{array}{ccc}0&\frac{f}{B_t}&0\\\frac{f}{B_t}&2(fU-|\omega|^2)&
fX^B-(\bome\xi^B+\omega\bxi^B)\\
0&fX^A-(\bome\xi^A+\omega\bxi^A)&-(\xi^A\bxi^B+\bxi^A\xi^B)
\end{array}\right). \nonumber \\
&&
\end{eqnarray}
In addition, we shall use the same gauge condition \meq{gaugec}.

Based on this ansatz, the action function $I$ of an outgoing
particle should satisfy following equation :
\begin{eqnarray}
g^{\mu\nu}\partial_{\mu}I\partial_{\nu}I+m^2=0.\label{I1}
\end{eqnarray}

Comparing with null geodesic method, we also need to control other
components of metric near the horizon. Note that the gauge choice
in section \ref{geometry} has fixed the metric components on
$\cH$. The behavior of metric components near the horizon is
controlled by Cartan structure equations,
\begin{eqnarray}
&&\frac{\partial U}{\partial
r}=(\vps+\bvps)-\frac{\bpi\bome+\pi\omega}{f}+D\ln f,\nonumber\\
&&\frac{\partial X}{\partial r}=-\frac{\bpi\bxi^4+\pi\xi^3}{f},\nonumber\\
&&\frac{\partial\omega}{\partial
r}=\bpi-\frac{\blam\bome}{f}-\frac{\mu\omega}{f}+\de\ln f,\nonumber\\
&&\frac{\partial\xi^3}{\partial r}=-\frac{\blam\bxi^4+\mu\xi^3}{f},\nonumber\\
&&\frac{\partial\xi^4}{\partial
r}=-\frac{\blam\bxi^3+\mu\xi^4}{f},
\end{eqnarray}
where $D:=l^a\del_a$, $\de:=m^a\del_{a}$. Here we have used the
modified Bondi coordinates $(t,r,\theta,\phi)$. The behavior of
metric near horizon is
\begin{eqnarray}
&&U=(\vps+\bvps)r+o(r),\nonumber\\
&&X=X_0+O(r),\nonumber\\
&&\xi^3=O(1),\nonumber\\
&&\xi^4=O(1),\nonumber\\
&&\omega=O(r).\label{asymTe}
\end{eqnarray}
The time derivatives of these quantities are
\begin{eqnarray}
\frac{\partial U}{\partial t}=o(r),\nonumber\\
\frac{\partial X}{\partial t}=O(r),\nonumber\\
\frac{\partial\omega}{\partial t}=O(r),\nonumber\\
\frac{\partial X^A}{\partial t}=O(r).\label{asymTed}
\end{eqnarray}

We consider the variable separation solution for $I$
\begin{eqnarray}
I=V(t)+W(r)+J(x^A).
\end{eqnarray}
Eq.(\ref{I1}) then becomes
\begin{eqnarray}
&&2(fU-|\omega|^2)(W')^2\nonumber\\
&&+2[(fX^A-\bome\xi^A-\omega\bxi^A)J_A+\frac{f}{B_t}\dV]W'\nonumber\\
&&-(\xi^A\bxi^B+\bxi^A\xi^B)J_AJ_B+m^2=0,\label{W}
\end{eqnarray}
where `` $\cdot$ " means $\frac{\partial}{\partial t}$, `` ' " means
$\ppr$ and $J_A=\frac{\partial J}{\partial x^A}$. Consequently, we
have
\begin{eqnarray}
W(r)&=&\int\frac{-\cB+\sqrt{\cB^2-\cA\cC}}{\cA}\ dr,\label{W} \eean
where \bean
\cA&=&2(fU-|\omega|^2),\nonumber\\
\cB&=&(fX^A-\bome\xi^A-\omega\bxi^A)J_A+\frac{f}{B_t}\dV,\nonumber\\
\cC&=&-(\xi^A\bxi^B+\bxi^A\xi^B)J_AJ_B+m^2.\nonumber
\end{eqnarray}
Taking the time derivative on both sides of \eq{W} gives
\begin{eqnarray}
&&2({\dot f}U+f{\dot
U}-\bome{\dot\omega}-\omega{\dot\bome})(W')^2\nonumber\\
&&+2\left[({\dot f}X^A+f{\dot
X}^A-{\dot\bome}\xi^A-\bome{\dot\xi}^A-{\dot\omega}\bxi^A-\omega{\dot\bxi}{}^A)J_A\right.\nonumber\\
&&\qquad\left.+\frac{{\dot f}}{B_t}\dV+\frac{f}{B_t}{\ddot
V}\right]W'\nonumber\\
&&-\frac{\partial}{\partial
t}(\xi^A\bxi^B+\bxi^A\xi^B)J_AJ_B=0,\label{W2}
\end{eqnarray}
Because \eq{W2} holds smoothly in the neighborhood of $\cH$, with
the help of the asymptotic behavior of the tetrad (\ref{asymTe}) and
(\ref{asymTed}), we get ${\dot X}^AJ_A+\frac{\ddot V}{B_t}=0$.
 Furthermore,
similar derivation shows $\partial_B(X^AJ_A)=0$, i.e.
$X^AJ_A+\frac{\dot V}{B_t}\heq const.$ In the axisymmetric case,
$X_0$ is constant on $\cH$\cite{As04}. Thus, ${\dot V}$ and $J_A$
are constants. Comparing with the stationary case, $E=-{\dot V}$
is the energy of this particle and $J_A$ is the angular momentum
of the particle. The energy condition guarantees $E-B_tX_0^AJ_A\ge
0$. With the above results and the asymptotic behavior, we find
that integral (\ref{W}) has a simple pole at horizon. The
imaginary part of the action function is determined by this pole.
Specifically,
\begin{eqnarray}
Im(I)=Im\left(\int W' dr\right)=\pi\frac{E-B_tX_0^AJ_A}{\kappa_t},
\end{eqnarray}
where $\kappa_t=B_t(\vps+\bvps)$. Combining with the WKB
assumption $\Psi(x)=\exp(iI)$, we get the thermal spectrum
$\Gamma\sim\exp[-\be(E-B_tX_0^AJ_A)]$. As we emphasized in section
\ref{geometry}, Only a canonical time direction $t^a$ can lead to a
well-defined horizon mass and angular momentum, as well as the
black hole mechanical law.  For such an observer, $t^a\heq
B_tl^a-\Omega_t\psi^a$, where $\psi^a$ is a Killing vector of
$\cH$ and  tangent to the leaf $S$, $\Omega_t$ is the angular
velocity of the horizon. In this case, we choose $X^A_0$ as
$X^A_0=\Omega_t\psi^A$. Because of the conservation of total
energy and angular momentum, $E$ is the variation of the energy of
black hole and $\psi(J)$ is the variation of the black hole.
Combining with the quasi-local black hole mechanical law
(\ref{1}), we obtain $\Gamma\sim\exp(\De\frac{a_{\cH}}{4})$.

\subsection{Freedom of coordinates}\label{sec-coo}
The calculation in section \ref{mone} appears to rely on a
specific coordinate system. The purpose of this subsection is to
investigate how much the result depends on the choice of
coordinates. In order to calculate the tunnelling effect, the
crucial step is to calculate the residue of $1/\dot r$. We
consider the coordinates $(t,r,\theta,\phi)$ as defined in section
\ref{geometry}, where $t$ is the canonical time. It will be
sufficient to focus on the following ine element
\bean ds^2=g_{tt} dt^2+2g_{tr} dtdr+ g_{rr}dr^2+g_{AB}dx^Adx^B.
\label{metric}
\eean
 These coordinates are nonsingular through the
horizon. By letting $ds^2=0$,  $\dot r$, the derivative of $r$
along an outgoing null geodesic, can be easily solved as
\bean \dot r=\frac{-g_{tr}+\sqrt{g_{tr}^2-g_{rr}(g_{tt}+g_{AB}{\dot
x}^A{\dot x}^B)}}{g_{rr}} \,. \label{rdw}
\eean
 Since $\ppa{t}^a $
approaches the horizon, $g_{tt}$ must vanish on the horizon. Since
$r$ is chosen to be a constant on the horizon, $\dot r$ also
vanishes on the horizon. Then we have
\bean g_{rr}(g_{tt}+ g_{AB}{\dot x}^A{\dot x}^B) \heq 0\,.
\label{voh}
\eean
Note that $g_{tt}\heq 0$ and the metric is non-degenerate around
the horizon. Thus $g_{tr}$ must be non-zero at the horizon.  So we
can simplify \eq{rdw} by Taylor expanding the square root term
around $g_{tr}^2$.  Then we find
\bean \frac{1}{\dot r}\approx-\frac{2g_{tr}}{g_{tt}+ g_{AB}{\dot
x}^A{\dot x}^B} \label{ordw} \eean It is not difficult to see that
the approximation made in \eq{ordw}  preserves the residue. Then
 \eq{ordw} leads to the desired emission rate as shown in section
\ref{mone}. Now consider the coordinate transformation
\bean
t&=&T+f(R)   \nonumber \\
r&=&R \label{ctr}\,, \eean where $f(R)$ is a smooth function
around the horizon. Under the new coordinates, the metric becomes
\bean
ds^2&=&g_{tt}dT^2+[2g_{tr}+g_{tt}f'(r)]dTdR\nonumber\\
&&+[g_{rr}+2g_{tr}f'(r)+g_{tt}f'(r)^2]dR^2\nonumber\\
&&+g_{AB}dx^Adx^B, \label{newm}
\eean
 which gives the radial null
geodesic equation
\bean
&&[g_{rr}+2g_{tr}f'(r)+g_{tt}f'(r)^2]\dot
R^2\nonumber\\
&&+[2g_{tr}+g_{tt}f'(r)]\dot R+g_{tt}+ g_{AB}{\dot x}^A{\dot
x}^B=0\,,
\eean
 where $\dot R=dR/dT$. The outgoing solution for
$\frac{1}{\dot R}$ is
\begin{widetext}
\bean \frac{1}{\dot R}=\frac{
2\left(g_{rr}+2g_{tr}f'(r)+g_{tt}f'(r)^2\right)}{-2g_{tr}-g_{tt}f'(r)
+\sqrt{(2g_{tr}+g_{tt}f'(r))^2-4[g_{rr}+2g_{tr}f'(r)+g_{tt}f'(r)^2][g_{tt}+
g_{AB}{\dot x}^A{\dot x}^B] }} \label{ord}\,.\nonumber\\ \eean
\end{widetext}
 Note that $g_{tt}=0$ at the horizon. Since the
coordinates are nonsingular,  $g_{tr}$ cannot be zero at the
horizon. In order to find the residue of $1/\dot R$, we Taylor
expand the square root around $(2g_{tr}+g_{tt}f'(r))^2$ to order
one (terms with higher orders are not relevant to the calculation
of residue). Then
\eq{ord} becomes
\bean \frac{1}{\dot R}\approx-\frac{2g_{tr}+g_{tt}f'(r)}{g_{tt}+
g_{AB}{\dot x}^A{\dot x}^B}  \,. \label{te} \eean Noting that
$g_{tt}$ vanishes on the horizon and comparing \eq{te} with
\eq{ordw}, we see that $1/\dot R$ has the same residue as $1/\dot
r$.

Now we are going to show that the result is invariant under a
rescaling of $r$. Consider the following coordinate
transformation:
\bean
t&=&\tilde t \nonumber \\
r&=&h(\tr)\,, \label{cor} \eean where $\tr$ is any smooth function
of $r$ satisfying $\tr=0$ on the horizon. Similarly, we find
\bean
\frac{1}{\dot{\tr}}&=&\frac{f'(\tr)}{-\frac{g_{tr}}{g_{rr}}+\sqrt{
\left(\frac{g_{tr}}{g_{rr}}\right)^2-\frac{g_{tt}+ g_{AB}{\dot
x}^A{\dot x}^B}{g_{rr}}}}
\nonumber \\
&\approx &\frac{2g_{tr}f'(\tr)}{g_{tt}+ g_{AB}{\dot x}^A{\dot
x}^B}\,. \eean Suppose the residue of $1/\dot r$ at the horizon is
$\alpha$. Then
\bean
\frac{1}{\dot{\tr}}\rightarrow \frac{\alpha f'(\tr)}{r}=\frac{\alpha
f'(\tr)}{f(\tr)}\approx\frac{\alpha
f'(\tr=0)}{f'(\tr=0)\tr}=\frac{\alpha}{\tr}\,,\label{rts}
\eean
which shows that $1/\dot r$ and $1/\dot{\tr}$ have the same
residue on the horizon.

We have shown that there exist two classes of coordinate
transformations, \eq{ctr} and \eq{cor}, that keep the residue
invariant in the tunnelling effect. However, the two
transformations share a common feature: they both preserve the
tangent field $\left(\frac{\partial}{\partial t}\right)^a$. So we
may conclude that the emission rate is determined only by a
specific family of orbits approaching the horizon and
parameterized by some coordinate time $t$ (which is the Killing
time in the Schwarzschild case). All comoving coordinate systems
preserving the tangent field $\left(\frac{\partial}{\partial
t}\right)^a$ are equivalent in the sense that they lead to the
same emission rate.

Now we keep the orbits fixed on the horizon but vary the orbits
off the horizon. This can be realized by the following coordinate
transformation
\bean
r&=&h(T,R) \nonumber \\
t&=&T \label{htr}\,.
\eean
where $\dot h=\frac{\partial h}{\partial T}\heq 0$. Then
\bean
\ppa{T}^a=\ppa{t}^a+\dot h \ppa{r}^a \,,\label{ptpt}
\eean
which shows that $\ppa{t}^a$ and $\ppa{T}^a$ coincide on the
horizon but are different off the horizon. By a similar
derivation, we have
\bean
\frac{1}{\dot R}&=&-\frac{g_{rr}h'}{g_{tr}+g_{rr}\dot
h-\sqrt{g_{tr}^2-g_{rr}(g_{tt}+ g_{AB}{\dot x}^A{\dot x}^B)}} \nonumber\\
 &\approx& \frac{-2 g_{tr}h'}{2g_{tr}\dot h+g_{tt}+ g_{AB}{\dot x}^A{\dot x}^B} \,.
\label{irh}
\eean
 A derivation similar to \eq{rts} shows that $h'$
in \eq{irh} does not change the residue. However, if $\dot h\sim
r$ near the horizon, it will cause a change of residue. For $\dot
h\sim r^n$ ($n\geq 2$), the residue keeps invariant. Since $\dot
h$ measures the difference of the two families of orbits
approaching the horizon (see \eq{ptpt}), the above results show
that when the difference is in the order of $r$, the residue will
change. But if the difference is smaller than $r$ (higher order of
$r$), the residue is unchanged. Thus, if the two families of
orbits in \eq{ptpt} satisfy $\dot h\sim r^n$, $n\geq 2$, we call
them equivalent in the sense that they lead to the same residue.

\section{Discussion}\label{conclusion}
The Hawking-like radiation has been derived as a tunneling process
near weakly isolated horizons. The null geodesic method and
Hamilton-Jacobi ansatz lead to the same result. This indicates
that thermal radiation is a generic property of horizon, not only
for stationary black holes. The choice of time direction has
played a key role in both methods. However, there are subtle
differences in the two methods. In the null geodesic method, only
the canonical time direction can define the horizon mass and lead
to the first law of black hole mechanics. In the Hamilton-Jacobi
method, the thermal spectrum exists for any choice of time
direction. However, only when the canonical time direction is
chosen, can the $\kappa_t$ term in the expression be interpreted
as the Hawking temperature. We also find that different observers
will give different temperatures even if they have the same limit
curve at the horizon. We find the difference
$\partial_t-\partial_{t_c}\sim O(r^2)$ can insure the invariance
of the temperature, where $t_c$ is the canonical time given by
Ashtekar {\em et al.}\cite{As04}. This result is reasonable
because the energy and angular momentum measured by arbitrary
observers cannot be directly related to the horizon mass. The
canonical time direction corresponds to the Killing observers in
Schwarzschild spacetime, where other observers are like the
Rindler observers of the Unruh effect.

Our discussion in this paper is confined to vacuum solutions.
There is no obvious difficulty to extend the discussion to matter
fields, for example, the Maxwell field or other gauge fields. When
matter is considered, the field equations can be controlled in
similar ways and the first law including mater fields is also
known\cite{Hay94}-\cite{As04}. The next interesting thing is
studying dynamical horizons. Some special cases have been
investigated\cite{zhao061}. The general calculation on dynamical
horizons will be done in future work.

\section*{Acknowledgements}
X.Wu is supported by NSFC Grant No. 10375087, Post-doctor
foundation of China and K.C.Wong Education Foundation, Hong Kong.
S.Gao is supported by NSFC Grants 10605006,  10373003, and the
Scientific Research Foundation for the Returned Overseas Chinese
Scholars, State Education Ministry. The authors would like to
thank Prof.Z.Zhao, Prof.C.G.Huang and Prof.Y.Tian for their
helpful discussions.

\end{document}